\documentclass{PoS}

\def\beq{\begin{equation}}
\def\eeq{\end{equation}}
\def\ber{\begin{eqnarray}}
\def\eer{\end{eqnarray}}
\def\bdm{\begin{displaymath}}
\def\edm{\end{displaymath}}

\def\mc{\mathcal}
\def\pa{\partial}
\def\d{\delta}

\def\th{\theta}

\title{Thirring Model with Jump Defect}

\ShortTitle{Thirring Model with Jump Defect}

\author{\speaker{A.R. Aguirre}\\
        Instituto de F\'{\i}sica Te\'orica, UNESP, S\~ao Paulo, Brazil.\\
        E-mail: \email{aleroagu@ift.unesp.br}}

\author{J.F. Gomes\\
       Instituto de F\'{\i}sica Te\'orica, UNESP, S\~ao Paulo, Brazil.\\
       E-mail: \email{jfg@ift.unesp.br}}

\author{L.H. Ymai\\
       Instituto de F\'{\i}sica Te\'orica, UNESP, S\~ao Paulo, Brazil.\\
       E-mail: \email{leandroy@ift.unesp.br}}

\author{A.H. Zimerman\\
       Instituto de F\'{\i}sica Te\'orica, UNESP, S\~ao Paulo, Brazil.\\
       E-mail: \email{zimerman@ift.unesp.br}}

\abstract{The purpose of our  work is to  extend the formulation  of classical affine Toda Models in the presence of jump defects to pure
fermionic Thirring  model.  As a first attempt 
 we construct the Lagrangian of the Grassmanian Thirring model with jump defect  (of Backlund type) and 
present its conserved modified momentum and energy expressions giving a first indication  of its integrability.}

\FullConference{5th International School on Field Theory and Gravitation,\\
		 April 20 - 24 2009\\
		 Cuiab\'a city, Brazil}

\begin{document}

\section{Introduction}
Recently, there has been some interest in the question of whether relativistic integrable field theories admit
 certain discontinuities (or jump defects) and, if so, what kind of properties they might have \cite{Zam,Cor,Cor1,Cor2,Cor3}. In particular, in \cite{Zam,Cor,Cor1,Cor2,Cor3} using the Lagrangian formulation, it was noted that
some bosonic field theories could allow discontinuities like  jump defects and yet remain classically integrable. The best understood of these models is the sinh-Gordon in which different soliton solutions are connected  in such way that the integrability is preserved.\\
The defect conditions relating fields evaluated as limits from both sides of the  defect turn out to be described by 
B\"acklund transformation located at the defect. This fact sheds an interesting new light on the role which the B\"acklund transformations have played in the development of soliton theory.

More recently, the extension to classical integrable $N=1$ and $N=2$ super sinh-Gordon models  involving bosons and fermions admitting jump defects have been studied \cite{Lea1,Lea2} using both the Lagrangian approach and the  zero curvature formalism.  
Here we study the Thirring model with jump defects as an example  of pure fermionic case.
In this work we explicitly construct its  modified conserved momentum and energy which gives a very strong suggestion of its classical integrability.

\section{Lagrangian Approach}

The Lagrangian density for Grassmanian Thirring model with  jump defect can be written as follows,
\begin{eqnarray}
 \cal{L} &=& \th(-x)\mc{L}_1 + \th(x)\mc{L}_2 +\d(x) \mc{L}_D \,,
\end{eqnarray}
where
\begin{eqnarray}
 \mc{L}_p &=& \frac{i}{2}\psi_1^{(p)}(\pa_t - \pa_x)\psi_1^{\dagger(p)} +\frac{i}{2}\psi_1^{\dagger(p)}
 (\pa_t - \pa_x)\psi_1^{(p)} +\frac{i}{2}\psi_2^{(p)}(\pa_t + \pa_x)\psi_2^{\dagger(p)}+\frac{i}{2}\psi_2^{\dagger(p)}
 (\pa_t + \pa_x)\psi_2^{(p)}  \nonumber \\
 &+&  m\big(\psi_1^{(p)}\psi_2^{\dagger(p)} + \psi_2^{(p)}\psi_1^{\dagger(p)} \big) -
 g\big(\psi_1^{\dagger(p)}\psi_2^{\dagger(p)}\psi_2^{(p)}\psi_1^{(p)}\big)\,,
\end{eqnarray}
is the lagrangian density for Thirring model \cite{Thi} describing massive two-component Dirac fields
$(\psi_1^{(p)},\psi_2^{(p)})$ with $p=1$ corresponding to $x<0$, $p=2$ corresponding to $x >0$, and $g$ is
coupling constant. The defect contribution located at $x=0$ is described by
\begin{eqnarray}
 \mc{L}_D &=& \frac{1}{2}\Bigg[\frac{2ia}{m}X^{\dagger}\pa_t X + i\psi_1^{(1)}\psi_1^{\dagger(2)} +i\psi_2^{(1)}\psi_2^{\dagger(2)} -i\psi_1^{(2)}
 \psi_1^{\dagger(1)} -i\psi_2^{(2)}\psi_2^{\dagger(1)}+\nonumber \\&+&\left(i(\psi_1^{(2)}-\psi_1^{(1)}) + a(\psi_2^{(2)}+
 \psi_2^{(1)}) \right)X^{\dagger} + \left(i(\psi_1^{\dagger(2)}-\psi_1^{\dagger(1)}) - a(\psi_2^{\dagger(2)}+
 \psi_2^{\dagger(1)}) \right) X \nonumber \\
 &-&\frac{iag}{2m}\left(i(\psi_1^{\dagger(2)}-\psi_1^{\dagger(1)})\psi_1^{(2)}\psi_1^{(1)} +
 a^{-1}(\psi_2^{\dagger(2)}+\psi_2^{\dagger(1)})\psi_2^{(2)}\psi_2^{(1)}\right) X^{\dagger}  \nonumber \\
 &-&\frac{iag}{2m}\left(-i\psi_1^{\dagger(1)}\psi_1^{\dagger(2)}(\psi_1^{(2)}-\psi_1^{(1)}) +
 a^{-1}\psi_2^{\dagger(1)}\psi_2^{\dagger(2)}(\psi_2^{(2)} +\psi_2^{(1)}) \right) X \nonumber \\
 &+& \frac{ag}{m}\psi_1^{\dagger(1)}\psi_1^{\dagger(2)}
 \psi_1^{(2)}\psi_1^{(1)} + \frac{g}{am}\psi_2^{\dagger(1)}\psi_2^{\dagger(2)}\psi_2^{(2)}\psi_2^{(1)}\Bigg]
\end{eqnarray}
Here, we introduce  auxiliary fields $X$ and $X^{\dagger}$.  

The field equations for $x\neq 0$ are obtained as
\begin{eqnarray}
 i(\pa_t -\pa_x) \psi_1^{(p)} &=& m \psi_2^{(p)} + g\psi_2^{\dagger(p)}\psi_2^{(p)}\psi_1^{(p)}\,,\label{e1}\\
 i(\pa_t + \pa_x) \psi_2^{(p)} &=& m\psi_1^{(p)} + g\psi_1^{\dagger(p)}\psi_1^{(p)}\psi_2^{(p)}\,,\label{e2}\\
 i(\pa_t -\pa_x) \psi_1^{\dagger(p)} &=& -m\psi_2^{\dagger(p)} -g\psi_1^{\dagger(p)}\psi_2^{\dagger(p)}\psi_2^{(p)}\,,\label{e3}\\
 i(\pa_t + \pa_x) \psi_2^{\dagger(p)} &=& -m\psi_1^{\dagger(p)} - g\psi_2^{\dagger(p)}\psi_1^{\dagger(p)}\psi_1^{(p)}\,,\label{e4}
\end{eqnarray}
which are the equations of motion for the Thirring model.

For $x=0$, the equations corresponding to defect conditions  are
\begin{eqnarray}
 X &=& (\psi_1^{(2)} + \psi_1^{(1)}) + \frac{iag}{2m}(\psi_1^{\dagger(2)} + \psi_1^{\dagger(1)})\psi_1^{(2)}
 \psi_1^{(1)}\nonumber\\&=& ia^{-1}(\psi_2^{(2)} - \psi_2^{(1)}) -\frac{g}{2a^2 m}(\psi_2^{\dagger(2)} -\psi_2^{\dagger(1)})
 \psi_2^{(2)}\psi_2^{(1)}  \label{b1} 
\end{eqnarray}
and its hermitian conjugated
\begin{eqnarray}
 X^{\dagger} &=& (\psi_1^{\dagger(2) } + \psi_1^{\dagger(1)}) - \frac{iag}{2m}(\psi_1^{(2)} + \psi_1^{(1)})\psi_1^{\dagger(1)}
 \psi_1^{\dagger(2)}\nonumber\\&=&-ia^{-1}(\psi_2^{\dagger(2)} - \psi_2^{\dagger(1)}) -\frac{g}{2a^2 m}(\psi_2^{(2)} -\psi_2^{(1)})
 \psi_2^{\dagger(1)}\psi_2^{\dagger(2)}\label{b2}
\end{eqnarray}
together with
\begin{eqnarray}
\pa_t X &=& {{ma^{-1}}\over {2}} (\psi_1^{(2)} - \psi_1^{(1)}) - {{im}\over {2}} (\psi_2^{(2)} + \psi_2^{(1)})\nonumber \\
& - &{{g}\over {4}}\Big( i(\psi_1^{ \dagger (2)} - \psi_1^{ \dagger(1) })\psi_1^{(2)} \psi_1^{(1)} +a^{-1}  (\psi_2^{\dagger (2) } + \psi_2^{ \dagger(1) })\psi_2^{(2)} \psi_2^{(1)}\Big), \label{10}\\
\pa_t X^{\dagger}  &=& {{ma^{-1}}\over {2}} (\psi_1^{\dagger (2)} - \psi_1^{ \dagger(1) }) + {{im}\over {2}} (\psi_2^{ \dagger (2)} + \psi_2^{ \dagger(1) })\nonumber \\
& + &{{g}\over {4}}\Big( i \psi_1^{\dagger (1) } \psi_1^{\dagger(2)} (\psi_1^{(2)  } - \psi_1^{(1)}) -a^{-1} \psi_2^{ \dagger(1) } \psi_2^{ \dagger(2) } (\psi_2^{(2)} + \psi_2^{(1  )})\Big) \label{11}
\end{eqnarray}
Notice that we can also write
\begin{eqnarray}
X&=& (\psi_1^{(2)} + \psi_1^{(1)} ) + {{iag}\over {2m}}\psi_1^{(2)} \psi_1^{(1)} X^{\dagger} \nonumber \\
&=& (\psi_1^{(2)} + \psi_1^{(1)} ) + {{iag}\over {2m}}\psi_1^{(1)}  X^{\dagger} X \nonumber \\
&=&(\psi_1^{(2)} + \psi_1^{(1)} ) - {{iag}\over {2m}}\psi_1^{(2)}  X^{\dagger} X  \label{12}
\end{eqnarray}
and similarly for $X^{\dagger}$.
As a consequence  of (\ref{10}), (\ref{11}) and the Thirring equations of motion (\ref{e1}) - (\ref{e4}), we obtain
\begin{eqnarray}
 \big(\pa_t +\pa_x \big)X &=& ma^{-1}\big(\psi_1^{(2)} - \psi_1^{(1)}\big) - \frac{ig}{2}\big(\psi_1^{\dagger (2)}-
 \psi_1^{\dagger (1)} \big)\psi_1^{(2)}\psi_1^{(1)} \,,\label{b3} \\
 \big(\pa_t -\pa_x \big)X &=& -im\big( \psi_2^{(2)}+\psi_2^{(1)} \big) - \frac{g}{2a}\big(\psi_2^{\dagger(2)} +
 \psi_2^{\dagger (1)} \big)\psi_2^{(2)}\psi_2^{(1)}  \,. \label{b4}
\end{eqnarray}
which can be re-written as 
\begin{eqnarray}
 \big(\pa_t +\pa_x \big)X &=& ma^{-1}X -2ma^{-1} \psi_1^{(1)} - ig \psi_1^{ \dagger(1)} \psi_1^{(1)} X -ig X^{\dagger}X \psi_1^{(1)}
 \,,\label{c3} \\
 \big(\pa_t -\pa_x \big)X &=&  -maX -2mi \psi_2^{(1)} - ig \psi_2^{\dagger(1) } \psi_2^{(1)} X -g a X^{\dagger}X \psi_2^{(1)}\,. \label{c4}
\end{eqnarray}
The equations (\ref{b1},\ref{b2},\ref{12},\ref{b3},\ref{b4}) are the B\"acklund transformation for the classical anticommuting
Thirring model \cite{Ize}. In other words, in this approach the B\"acklund transformation at $x=0$
represents the boundary conditions between two regions.

\section{Conserved Momentum and Energy}

The canonical momentum (which is not expected to be preserved) is given by
\ber
 P &=& \int\limits_{-\infty}^{0}dx \left[\frac{i}{2}\Big(\psi_1^{(1)}\pa_x \psi_1^{\dagger (1)}+ \psi_1^{\dagger (1)}
 \pa_x \psi_1^{(1)} +\psi_2^{(1)}\pa_x \psi_2^{\dagger (1)} + \psi_2^{\dagger (1)}\pa_x \psi_2^{(1)}\Big) \right] \nonumber \\
 &\mbox{}& \!\!+\,\int\limits_{0}^{\infty}dx\left[\frac{i}{2}\Big(\psi_1^{(2)}\pa_x \psi_1^{\dagger (2)}+ \psi_1^{\dagger (2)}
 \pa_x \psi_1^{(2)} +\psi_2^{(2)}\pa_x \psi_2^{\dagger (2)} + \psi_2^{\dagger (2)}\pa_x \psi_2^{(2)}\Big) \right] .
\eer
Using the field equations (\ref{e1},\ref{e2},\ref{e3},\ref{e4}) we obtain
\begin{eqnarray}
\frac{dP}{dt} &=&
\left[m(\psi_1^{(1)}\psi_2^{\dagger(1)}+\psi_2^{(1)}\psi_1^{\dagger(1)}) -g\psi_1^{\dagger(1)}
\psi_2^{\dagger(1)}\psi_2^{(1)}\psi_1^{(1)}+\frac{i}{2}(\psi_1^{\dagger(1)}\pa_t\psi_1^{(1)}+\psi_1^{(1)}\pa_t
\psi_1^{\dagger(1)} \nonumber \right.\\ &\mbox{}& \left.+\,\psi_2^{\dagger(1)}\pa_t\psi_2^{(1)}+\psi_2^{(1)}\pa_t\psi_2^{\dagger(1)}) \right]_{x=0}
-\left[m(\psi_1^{(2)}\psi_2^{\dagger(2)}+\psi_2^{(2)}\psi_1^{\dagger(2)}) -g\psi_1^{\dagger(2)}
\psi_2^{\dagger(2)}\psi_2^{(2)}\psi_1^{(2)} \nonumber \right.\\ &\mbox{}& \left. +\frac{i}{2}(\psi_1^{\dagger(2)}\pa_t\psi_1^{(2)}+\psi_1^{(2)}\pa_t
\psi_1^{\dagger(2)}+ \psi_2^{\dagger(2)}\pa_t\psi_2^{(2)}+\psi_2^{(2)}\pa_t\psi_2^{\dagger(2)}) \right]_{x=0}
\end{eqnarray}\\
Considering the boundary conditions (\ref{b1},\ref{b2},\ref{b3},\ref{b4}) the right hand side becomes a total time derivative. Thus, we found a functional $P_D$ given by
\ber
 P_D &=& \frac{ia}{m}\big(X\pa_t X^{\dagger} - (\pa_t X)X^{\dagger}\big) - \frac{i}{2}\big(\psi_1^{(1)}
 \psi_1^{\dagger(2)}+\psi_1^{\dagger(1)}\psi_1^{(2)}+3\psi_2^{(1)}\psi_2^{\dagger(2)}+3\psi_2^{\dagger(1)}
 \psi_2^{(2)}\big) \nonumber \\
 &\mbox{}& -\, \frac{ga}{2m}\psi_1^{\dagger(1)}\psi_1^{\dagger(2)}\psi_1^{(2)}\psi_1^{(1)} + \frac{g}{2ma}\psi_2^{\dagger(1)}
 \psi_2^{\dagger(2)}\psi_2^{(2)}\psi_2^{(1)}\,,
\eer\\
so that ${\cal P}=P + P_D$ is conserved. This `modified' momentum $\cal{P}$  appears to be a `total' momentum which is
preserved containing bulk and defect contributions.
\noindent In the case of the energy
\ber
 E &=& \!\!\int\limits_{-\infty}^{0}dx \left[\frac{i}{2}\Big(\psi_1^{(1)}\pa_x \psi_1^{\dagger (1)}+ \psi_1^{\dagger (1)}
       \pa_x \psi_1^{(1)} -\psi_2^{(1)}\pa_x \psi_2^{\dagger (1)} - \psi_2^{\dagger (1)}\pa_x \psi_2^{(1)}\Big) \right. \nonumber\\
 &\mbox{}& \left. -m\big(  \psi_1^{(1)}\psi_2^{\dagger (1)} + \psi_2^{(1)}\psi_1^{\dagger (1)}\big) 
           + \, g\psi_1^{\dagger(1)}\psi_2^{\dagger (1)}\psi_2^{(1)}\psi_1^{(1)}\right] +
           \int\limits_{0}^{\infty}dx\,\,\big[(1) \leftrightarrow (2)\big] \, ,
\eer
the energy-like conserved quantity is $\mc{E} = E + E_D $ , where
\ber
 E_D &=& \frac{ia}{m}\big(X\pa_t X^{\dagger} - (\pa_t X)X^{\dagger} \big) - \frac{i}{2}\big(\psi_1^{(1)}\psi_1^{\dagger(2)}
 +\psi_1^{\dagger(1)}\psi_1^{(2)} +\psi_2^{(1)}\psi_2^{\dagger(2)}+\psi_2^{\dagger(1)}\psi_2^{(2)}\big) \nonumber \\
 &\mbox{}& -\,\frac{ga}{2m}\psi_1^{\dagger(1)}\psi_1^{\dagger(2)}\psi_1^{(2)}\psi_1^{(1)} - \frac{g}{2ma}\psi_2^{\dagger(1)}
 \psi_2^{\dagger(2)}\psi_2^{(2)}\psi_2^{(1)}\,.
\eer
We notice that this functional $E_D$ appears on-shell to be the defect lagrangian ${\cal L}_D$. This property already was found in the sinh-Gordon model with defect \cite{Cor}.
 Then, in spite of the loss of translation invariance, the fields can exchange both energy and momentum with the defect.


In this work we have studied the classical integrability of the Grassmanian Thirring model
with jump defect by constructing the lowest conserved quantities, namely, the modified momentum and energy.  The integrability of the model involves also higher conservation laws  which are encoded within the Lax pair formalism.  This work is in progress.

\acknowledgments
We would like to thank the organisers of the FIFTH INTERNATIONAL SCHOOL
ON FIELD THEORY AND GRAVITATION for the opportunity to present these ideas. 
ARA and LHY thank  FAPESP,  AHZ and JFG   CNPq for financial support.  

\end{document}